\documentclass[aps,prb,twocolumn,groupedaddress,amsmath,amssymb,showpacs]{revtex4}

\usepackage{graphicx}
\usepackage{dcolumn}
\usepackage{bm}

\newcommand{\be}{\begin{equation}} 
\newcommand{\ee}{\end{equation}}
\newcommand{\ba}{\begin{eqnarray}}
\newcommand{\ban}{\begin{eqnarray*}}
\newcommand{\ean}{\end{eqnarray*}}
\newcommand{\ea}{\end{eqnarray}}
\newcommand{\Li}{$\textrm{LiHoF}_\textrm{4}$} 
\newcommand{\Ho}{$\textrm{Ho}^{3+}$\ }
\DeclareMathOperator{\sgn}{sgn}
\begin{document}
\title{Absence of domain wall roughening in a transverse-field Ising
    model with long-range interactions}
\author{George I. Mias}
\email{george.mias@yale.edu}
\author{S. M. Girvin}
\email{steven.girvin@yale.edu}
\affiliation{Sloane Physics Laboratory,Yale University, New Haven, CT
06520-8120}
\date{\today}
\begin{abstract} We investigate roughening transitions in the context of 
transverse-field Ising models. As a modification of the transverse Ising model
with short range interactions, which has been shown to exhibit domain wall
roughening, we have looked into the possibility of a roughening transition for
the case of long-range interactions, since such a system is physically realized
in the insulator LiHoF$_4$.  The combination of strong Ising anisotropy and
long-range forces lead naturally to the formation of domain walls but we find
that the long-range forces destroy the roughening transition.
\end{abstract}
\maketitle
\section{\label{Introduction} Introduction}
Magnetic systems present us with the opportunity to study not only classical but
also quantum critical phenomena and thus provide us with unique insights in
condensed matter physics and the intertwining of classical and quantum
mechanics\cite{Subir}.  Of particular interest is the transverse field Ising
model\cite{Subir, bitko, prabuddha}, whose representation in terms of Pauli spin
matrices is
\be H=\sum^N_{i,j}
J_{ij}\hat{\sigma}_i^z\hat{\sigma}_j^z-h_x\sum_i^N\hat{\sigma}_i^x%
,\label{transIsing}
\ee where $h_x$ represents an applied transverse magnetic field and $J_{ij}$ are
coupling constants. Notice that in the absence of the magnetic field the
Hamiltonian is diagonal in the $\hat{\sigma^z}$ basis, and the system is simply
a classical Ising model.  As the magnetic field is turned on, $\hat{\sigma}^x$
operators are introduced which do not commute with the $\hat{\sigma^z}$
operators. Thus, turning the transverse field on or off essentially turns
quantum mechanics on or off in the system.  The spins of this system align at
temperature T=0 in a  ferromagnetic ground state \cite{Subir} whereas  at high
temperatures the system becomes disordered. In the absence of a transverse
magnetic field, a continuous phase transition occurs between the paramagnetic
and ordered ferromagnetic states. This phase transition is driven by thermal
fluctuations.

Applying a transverse magnetic field, $\vec{h}_x$, perpendicular to the axis of
preferred magnetization, can also cause a transition between the ferromagnetic
and  the disordered states even at zero temperature \cite{Subir, bitko}.  This
behavior is driven by quantum zero-point fluctuations of the z component of the
spins due to the transverse field.

We now imagine the system described by (1) has domain walls - this is achieved
in principle by imposing appropriate anti-periodic boundary conditions.  It is
expected that, in the absence of the transverse field, at T=0 the domain walls
would be flat.  As we increase temperature we would first observe nucleation of
steps in the interface.  At some temperature, called the roughening temperature
$T_R$,  the entropy of thermal fluctuations of the interface would dominate the
interfacial energy and the interface would become rough\cite{lapujoulade}.  At
temperatures above the roughening transition, the amplitude of fluctuations in
the surface's position scales as L$^\theta$ where L is the system size and 
$\theta$ is an exponent characterizing the nature of the interface.  Thus the
interface fluctuations would diverge as the system length increases to infinity.
 However, we note that if $\theta < 1$ then the system size would diverge faster
than the fluctuations and so the fluctuations of the interface would not
overwhelm the bulk order in the system.  This would allow for a roughening
transition to take place at a temperature lower than the bulk order-disorder
phase transition temperature.

\begin{figure}[tb]
\includegraphics*{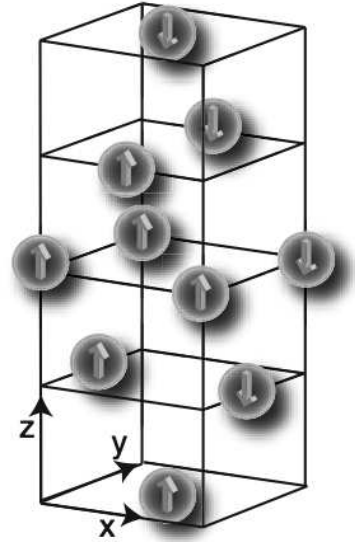}
\caption{\label{latticediagram} \Li\ behaves essentially as an Ising model with
long range interactions.  Interesting physics arises with the application of a
transverse magnetic field.}
\end{figure}

A physical realization of the transverse Ising model as described by
Eq.~(\ref{transIsing}) is  provided by the insulating magnet \cite{bitko, prabuddha}
\Li.  In this case, the coupling constants  $J_{ij}$ are actually no longer
near-neighbor but dipolar in nature and so decay as the inverse distance cubed,
$\sim \frac{1}{r^3}$.  In \Li\ the \Ho ions are responsible for the magnetic
behavior. The crystal field splitting of the \Ho states is such that the ground
state is doubly degenerate and well below the higher states, thus leading to a
very strong Ising anisotropy.  The two spin states for each \Ho ion point along
the z-axis, as shown in Fig.~\ref{latticediagram}.  The phase diagram, Fig.~\ref{phases}, is indicative of a quantum phase transition at T = 0 and at a
critical transverse magnetic field\cite{bitko} { $\vec{H}_x = \vec{H}_c \ \sim$
49 kOe, as well as a classical phase transition between ferromagnetic and
disordered states in the absence of a magnetic field at T$_c$=1.53 K.

The dipolar long-range forces and the insulator nature of \Li\ lead to the
natural formation of needle-like alternating domains of antiparallel
magnetization, without the need to force antiperiodic boundary conditions upon
an experimental sample.  Since \Li\ is an insulator, the electrons and hence the
constituent spins of the system are localized.  The large anisotropy of the
system means the domain walls which separate regions of opposite magnetization
in a \Li\ sample are sharp and well defined, in the sense that as we move across
a domain interface the system abruptly switches from one magnetization state to
another.  This is to be contrasted to domain walls found in metals which are
extended and provide for a continuous smoother transition of the magnetization
as we move from one domain to the next.  Thus, the first impression we get is
that \Li\ appears as an ideal system to study roughening transitions of domain
interfaces.  However, our analysis suggests otherwise.

In what folllows we first briefly review the case of only short-ranged exchange
interactions being present, where we would expect to see a roughening
transition, as described by Fradkin\cite{fradkin83}.   Then, we demonstrate that
in the case of dipolar forces such as seen in \Li\ the long-range nature of the
interactions drives the roughening temperature up to the bulk transition
temperature, so that the domain walls remain flat throughout the ferromagnetic
regime.  We show how to obtain a Hamiltonian for a field theoretical description
of a single domain wall, which results in a modified sine-Gordon model that
behaves as the regular sine-Gordon model but with effectively higher
dimensionality.  We see how this behavior is verified by a renormalization group
calculation using a smooth cutoff  and conclude that the effective higher
dimensionality of the domain wall, d$>$2, indicates the absence of a roughening
transition and the persistence of flat domain walls in the ferromagnetic phase. 
Finally, we mention some further considerations for stepped interfaces.

\begin{figure}[tb]
\includegraphics[width=3.375in]{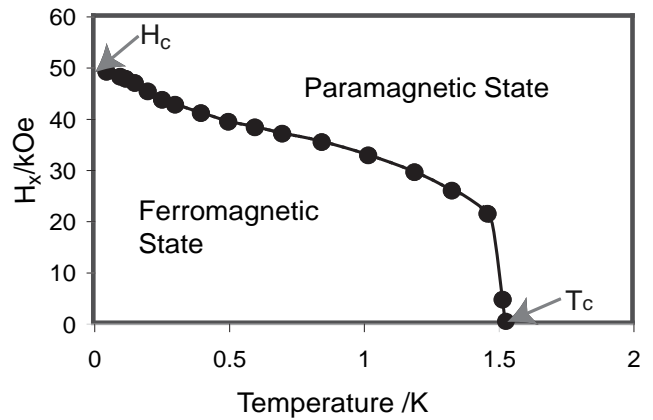}
\caption{\label{phases} \Li\ undergoes both a classical thermal phase transition
as well as a quantum phase transition with the application of a transverse
magnetic field, after Ref.~[\onlinecite{bitko}].}
\end{figure}

\section{\label{shortrange} Transverse Ising Model with Roughening: Short Range
interactions}

Using an Ising model for a two-state spin system with appropriate boundary
conditions, a two-region problem can be set up with an interface of finite width
separating regions of opposite spin.  Investigations of this interface for the
case of short-range interactions show that it undergoes the roughening
transition \cite{bufflovettstillinger,abrahamreed74,abraham81,
abraham82,abraham84,fisherweeks82}.  A numerical estimate for the roughening
transition was carried out some time ago by Weeks et al.\cite{weeksgilmerleamy}
for a 2-d interface in a classical Ising model, obtaining a roughening critical
temperature of $0.57T_c$, where $T_c$ is the bulk critical temperature for the
order-disorder transition.  It would be useful to redo this calculation with
higher precision using newer computational techniques.

In the case of two-dimensional interfaces it was proposed by Fisher and
Weeks\cite{fisherweeks83} that the interfaces of three-dimensional quantum
crystals are always smooth at zero temperature.  This suggests that there can be
no quantum roughening transition.  Fradkin\cite{fradkin83} investigated the
suggestion further and verified it for two different models, namely, a model
describing the solid to vacuum interface of a three-dimensional quantum crystal,
which is similar to the model used by Fisher and Weeks,  and a model describing
the interface of the three-dimensional quantum transverse Ising model. Fradkin
obtains the effective Hamiltonian of the two-dimensional interfaces, both for
the quantum crystal model,
\ba
H_{QC}&=&-K\sum_{\vec{r}}\cos[\hat{p}(\vec{r})-\hat{p}(\vec{r}~\!')]%
\nonumber%
\\
&+&\frac{J}{2}\sum_{<\vec{r},\vec{r}~\!'>}\left[n(\vec{r})-n(\vec{r}~\!'%
)\right]^2\label{qmcrystal},
\ea and for the transverse Ising model,
\ba H_{TI}&=&-K\sum_{\vec{r}}\cos\hat{p}(\vec{r})\nonumber\\
&+&\frac{J}{2}\sum_{<\vec{r},\vec{r}~\!'>}\left[n(\vec{r})-n(\vec{r}~\!'%
)\right]^2\label{transising},
\ea where $\hat{p}$ is canonically conjugate to $n$, the discrete height
variable.

Notice here the difference in the kinetic energy term which arises from the
difference in symmetry between the different models and gives rise to different
dynamics\cite{fradkin83}.  For the quantum crystal model the particle number
$N=\sum_{\vec{r}}n(\vec{r})$ is conserved while for the analogous quantity for
the transverse Ising model it is not.  Since our model for \Li\ is the
transverse Ising model we take into account the considerations by
Fradkin\cite{fradkin83} instead of the Fisher and Weeks
model\cite{fisherweeks83}.

The 2+1 dimensional quantum interface model for the transverse Ising is dual to
a three-dimensional general Coulomb gas which is known to have only a conducting
plasma phase\cite{kosterlitz77}.  Since the surface has to be smooth at T=0 this
implies that roughening can only occur if the bulk loses its long-range order,
so the critical magnetic field parameter for the roughening and order-disorder
transitions will coincide, i.e. $h_R=h_c$.

For finite temperatures, where the transition is classical, we can ignore the
cosine kinetic energy terms since these become unimportant,  and notice that the
two models in Eqs.~(\ref{qmcrystal}) and (\ref{transising}) become the same, namely a
realization of the discrete Gaussian SOS model \cite{fradkin83}.  A continuum
version of the discrete Gaussian SOS model is the sine-Gordon model
\cite{knops77}.  Briefly, the sine-Gordon model hamiltonian contains two terms:
\be H_{sG}=\int d^2\vec{r} \left[\frac{J}{2}\left(\nabla
n(\vec{r})\right)^2-y\cos\left(\frac{2\pi n(\vec{r})}{a}\right)\right]\label{sineGordon}.
\ee
The first term describes the cost of local fluctuations of the interface. 
The second term takes the effects of the lattice into account, and the fact that
the interface height field $n(\vec{r})$ can really only take discrete values,
namely multiples of the lattice spacing a.  This periodic term promotes the
pinning of the interface onto equidistant parallel planes, and thus encourages
the interface to remain smooth.  The sine-Gordon model undergoes a
Kosterlitz-Thouless phase transition\cite{kosterlitzthouless73,knops77} and
displays two distinct phases corresponding to the smooth surface at low
temperature, where the cosine term is perturbatively important, and a rough
phase at high temperatures, where the cosine term is unimportant. Thus we expect
a roughening phase transition at some finite temperature, $T_R$, less than the
bulk order-disorder transition temperature.  Given that the interface has to be
smooth at $T=0$ we propose the phase diagram shown in Fig.~\ref{3dphasediagram},
where we include, for completeness, the related renormalization group (RG)
flows.  For the two-dimensional interface we notice that in the temperature
region $T_R<T<T_c$ the flows are opposite for the two different phase
transitions.  This is related to the fact that the onset of high fluctuation for
the interface occurs at a temperature below the critical order-disorder
transition where the bulk system loses its long-range order.

\begin{figure}[tb]
\includegraphics[width=3.375in]{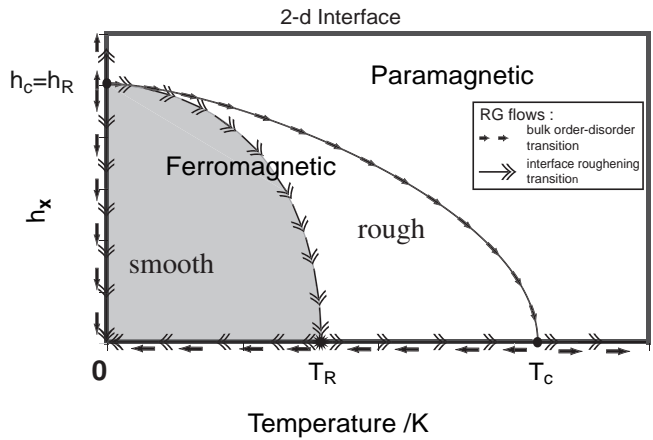}
\caption{\label{3dphasediagram} For a two-dimensional interface we expect a
thermal roughening transition at some finite temperature.  At zero temperature
the critical transverse field for the roughening transition coincides with that
of the bulk order-disorder transition.  The two sets of arrows indicate the
expected qualitative renormalization group flows for the different phase
transitions: the bulk order-disorder transition (plain arrows) and the interface
roughening transition (double-headed arrows).}
\end{figure}

It is also interesting to consider a one-dimensional interface in the transverse
field Ising model, in which case we expect the domain wall to be rough at any
finite temperatures.  However, as pointed out by Fradkin\cite{fradkin83}, at
zero temperature the transverse Ising model is dual to a two-dimensional Coulomb
gas\cite{chui76}. This is dual in the continuum limit to a  two-dimensional
sine-Gordon model and is known to have a metal-insulator phase transition
\cite{knopsouden,kosterlitzthouless73,kosterlitz77} at some value of $h_x$ at
T=0:  we expect a smooth interface for small $h_x$; as we increase $h_x$ we go
through a phase transition to a rough interface for some $h_R$, before the onset
of the order-disorder transition at $h_c$.  The proposed phase diagram and
qualitative RG flows are shown in Fig.~\ref{2dphasediagram}.  We see that for
the roughening transition, temperature is always a relevant parameter and flows
to a fixed point at infinity.

\begin{figure}[tb]
\includegraphics[width=3.375in]{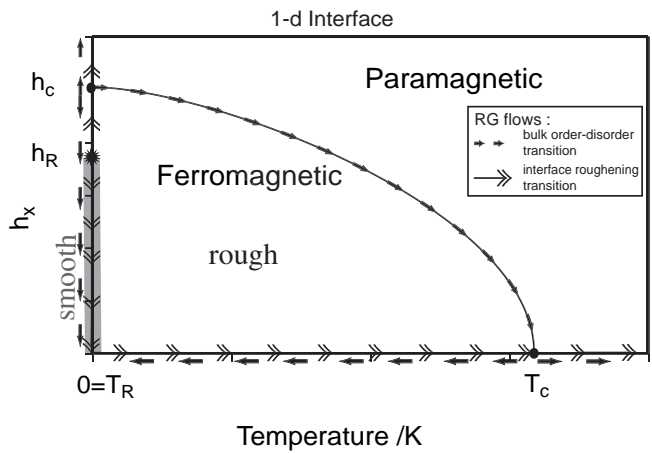}
\caption{\label{2dphasediagram} For a one-dimensional interface the system is
rough at all finite temperature.  The roughening transition occurs at zero
temperature.  In addition we expect a quantum roughening phase transition. The
two sets of arrows indicate the expected qualitative renormalization group flows
for the different phase transitions: the bulk order-disorder transition (plain
arrows) and the interface roughening transition (double-headed arrows).}
\end{figure}

\section{\label{macroscopics}\Li\ two-domain system:  Long-range Interactions}
As mentioned in the introduction, the behavior of \Li\ can be understood in
terms of a quantum Ising model in the presence of a transverse magnetic
field\cite{bitko,prabuddha}, in which the states of the \Ho ions are represented
by the Ising spins $|\uparrow\rangle$ and $|\downarrow\rangle$.  The effective
Hamiltonian looks like\cite{prabuddha}:
\ba H&=&\frac{1}{2}\sum_{i\ne j}J\frac{r_{ij}^2-3z_{ij}^2}{r_{ij}^5} S_i^zS_j^z
-h^x\sum_i S_i^x +\nonumber\\
&\ &+\frac{1}{2}\sum_{<i,j>}J_{ex}\sigma^z_i\sigma^z_j,  \label{tising}
\ea where $h^x$ corresponds to an applied transverse magnetic field and $J_{ex}$
is an effective antiferromagnetic exchange term that is added to obtain an
effective Hamiltonian that matches the experimental results.

Since \Li\ forms domains naturally, we consider for our calculations a system
consisting of two semi-infinite three-dimensional domains of antiparallel
magnetizations (directed in the positive or negative z-directions) that are
separated by a domain wall set at $x=0$.  We want to calculate the
configurational energy for fluctuations in the domain wall with respect to a
flat interface.   We proceed by assuming a height for the domain interface to be
a general function $\psi(y,z)$, so that the magnetization vector for the system
becomes,
\be
\vec{M}=\{0,0,m_0 \sgn(x-\psi(y,z)\}.
\ee We consider the classical magnetostatic problem and after we relate a
``magnetic charge density" $q(x,y,z)$ to the magnetization,
\be q(x,y,z)=-\nabla\cdot\vec{M},
\ee we see that in analogy with electrostatics, the extra energy associated with
having a non-flat profile is given in the continuum limit by\cite{landau8}
\be U=\frac{1}{2}\int_{V} q(\vec{r})\phi(\vec{r}),
\ee where $\phi$ is the magnetic potential given in three dimensions by
\be
\phi(\vec{r})=\int_{V'} \frac{q(\vec{r}~\!')}{|\vec{r}-\vec{r}~\!'|}.
\ee This results in the following expression for our system
\begin{widetext}
\be U=2 m_0^2\int dy dz dy' dz' \frac{ 
\partial_z \psi(y,z) 
\partial_{z'}\psi(y',z')}
{\left[\left(\psi(y,z)-\psi(y',z')\right)^2+(y-y')^2+(z-z')^2%
\right]^{1/2}}\label{ufull}.
\ee
\end{widetext}

We now assume that the profile of the interface does not vary greatly from one
position to another and so we can expand the denominator to zeroth order in
$\left[\psi(y,z)-\psi(y',z')\right]$, which is certainly true in the
``smoother''  regime of $\theta\ll1$ and probably asymptotically correct for all
$\theta<1$. We can rewrite the energy in this approximation in Fourier space to
obtain:
\be U_0 = 4 \pi m_0^2 \int \frac{dk_ydk_z}{(2\pi)^2}\frac{k_z^2}{|k|}
|\psi(k_y,k_z)|^2\label{u0}.
\ee Odd order terms vanish and higher even-order terms do not display the
singular behavior of the zeroth-ordered term which will ultimately dictate the
critical behavior of our system.

A point of concern here is that, in the continuum model used, the fluctuations
in the y-direction do not have any energy cost since they do not change the
angle between magnetization up and down in the z-direction as we move across the
interface. Thus we should consider fluctuations in the y-direction in a lattice
model rather than in the continuum to see their effect. A further complication
arises, since in addition to the dipolar Ising model interaction which would
raise the energy of the system there could be an effective exchange interaction
in \Li\, as indicated in Eq.~(\ref{tising}),  tending to lower the energy as the
domain wall fluctuates.  This effective exchange interaction has recently been
discussed by Chakraborty et al\cite{prabuddha}, who obtained an estimate of
$J_{ex}$  from  Monte Carlo simulations,
\be J_{ex}=0.75J=0.75\times0.214K.
\ee To get a feel for the system stability to deformations and the domain wall
surface tension we took a \Li\ system starting with two semi infinite domains of
anti-parallel magnetization (a cube with sides  extending to $\pm$ $N$) and then
we introduced a semi-infinite unit step deformation in the chosen y-direction. 
Briefly, this is a lattice sum calculation where the dipolar interactions were
summed over the infinite Bravais lattice, which for \Li\ has tetragonal unit
cells with dimensions $(a,a,c)=(1,1,2.077)a$, where   $a=5.175$\AA\ is the
lattice constant.  Each unit cell has four spins, at the positions
$\{0,0,0\},\{0,\frac{a}{2},\frac{c}{4}\},\{\frac{a}{2},\frac{a}{2},-\frac{c}{2%
}\}$ and $\{\frac{a}{2},0,-\frac{c}{4}\}$. The sums show slow convergence, $\sim
\frac{1}{x^3} $, so we used Ewald\cite{ewald} summation to obtain faster
convergence.  Our results indicate that the deformation causes an increase in
energy from the dipolar term of $\frac{1}{2}\times 2(2N+1)\times0.927J$ whereas
the exchange interaction would cause a decrease by $\frac{1}{2}\times
2(2N+1)\times 0.75J$. Thus, we have an overall positive energy cost to a single
step deformation and a positive surface tension associated with this
deformation.  Taking the zeroth order term of the Hamiltonian Eq.~(\ref{u0}) and
following the above discussion of our numerical considerations for the
deformation of an interface in equilibrium, we add an extra tension term, and
consider the free interface to be modelled by
\be U_0' = 4 \pi m_0^2 \int \frac{d^2\vec{k}}{(2\pi)^2}\left(\frac{k_z^2}{|k|}
+\gamma a k^2\right)|\psi(\vec{k})|^2.\label{u0tension}
\ee We note that at this Gaussian level the individual modes are not coupled and
hence we can evaluate the average of $U_0$ assigning a variance of
$\frac{1}{2\beta}$ to each mode. Thus we can write down the mean correlated
height difference, G, as
\ba G&=&\left<(\psi(y,z)-\psi(0,0))^2\right>\\
&=& \int \frac{d^2\vec{k}}{(2\pi)^2} \frac{1-e^{i \vec{k}\cdot\vec{r}}}{4\pi\beta
m_0^2\left( \frac{k_z^2}{|k|} +\gamma a k^2\right)}.\label{gshortrange}
\ea In the large distance limit, $G\sim O(L^ 0)$, thus indicating that the
interface has bounded fluctuations and is smooth at all temperatures.

\subsection{\label{terms} Long-range sine-Gordon model}

We now wish to investigate any possibility of critical behavior, by performing a
renormalization group analysis. Starting from the zeroth order term in our
modified energy expression Eq.~(\ref{u0tension}) we now include a sine-Gordon term
to take into account the existence of the lattice and write the full action as
\ba -\beta H_u&=&-\beta \Big{\{}4 \pi m_0^2 \int
\frac{d^2\vec{k}}{(2\pi)^2}\left(\frac{k_z^2}{|k|}+\gamma a k^2\right)
    \psi(\vec{k})\psi(\vec{-k})\nonumber \\
    &\ & - \frac{g_u}{a^2} \int d^2 r \cos\left(\frac{2
\pi\psi(\vec{r})}{a}\right)\Big{\}},
\ea where $\beta = k_BT$, $a$ is a lattice constant, and $\gamma$ and $g_u$ are
couplings for the surface tension term and fugacity of the system respectively. 
To make the coupling constants dimensionless we take $\beta m_0^2 \rightarrow
\frac{m^2}{a^3}, \beta g_u\rightarrow g $, and in addition we  define a
dimensionless field in real space, $\phi(r)=\frac{2\pi \psi(r)}{a},\
\phi(k)=\frac{2\pi}{a}\psi(k)$ so that our action becomes
\ba S &=& -\frac{m^2}{\pi}  \int \frac{d^2\vec{k}}{(2\pi)^2}
    \left(\frac{k_z^2}{a |k|} + \gamma k^2\right) \phi(\vec{k})\phi(-\vec{k})\nonumber\\
    &\ &+ \frac{g}{a^2} \int d^2 r \cos(\phi(\vec{r})). \label{s}
\ea The above action is just the familiar sine-Gordon field, with an extra
coefficient $k_z^2/a |k|$ in the quadratic part, which accounts for the long range
nature of the interactions. The sine-Gordon model can be renormalized via a
smooth cutoff approach \cite{knopsouden}. Note here that if we now try to
renormalize the above action using standard isotropic rescalings then the end
result is that the singular $\sim  \frac{k_z^2}{a|k|}$ term cannot be renormalized
and will dominate - since the renormalization of non-singular couplings cannot
generate singular terms.  Therefore, let us consider a different rescaling for
the x and y directions while,  as is usual, also requiring that the field
$\phi(\vec{r})$ does not change under rescaling so as to preserve the lattice
structure. We rescale so that $ z'=e^{-bs}z,\ y' = e^{-\alpha s}y,\ k_z' = e^{b
s}k_z,\ k_y' = e^{\alpha s} k_y,\ \phi'(\vec{k'}) = e^{-(b +\alpha)s}\phi(k),\
$and  $g'  =  e^{(b +\alpha)s}g,\ $ where $\alpha, b \geqslant 0. $

We concentrate on the terms quadratic in the field and choose a and b so that
the two terms in the quadratic prefactor scale the same way under the
transformation.  This cannot be done consistently unless we assume $\alpha < b$
which leads to the choice $\alpha = \frac{2}{3} b$.  We choose $\alpha = 1$ and
notice that terms which behave as $k_z^2$ become irrelevant in this RG scheme.

Bearing in mind the above discussion and in addition introducing a smooth
momentum cutoff function $f(k_y,k_z)$ in the simplified action to take the
effects of the lattice into account, we finally obtain the rescaled action
\ba S &=&  -\frac{m^2}{ \pi}  \int \frac{d^2\vec{k'} e^{\frac{s}{2}}}{(2\pi)^2
f(k_y,k_z)}
    \Big{\{}\frac{k_z'^2}{a |k_y'|}\nonumber \\
    &\ &+ \gamma (k_y'^2) \Big{\}}|\phi'(\vec{k}~\!')|^2
    +\frac{g e^{\frac{5s}{2}}}{a^2} \int d^2 r \cos\left(\phi(\vec{r})\right)\label{rescaledAction}
    \ea This has the form of the original hamiltonian if we define our new
couplings as:
\ba g'  &=& e^{\frac{5s}{2}}g,\\
m'^2    &=& e^{\frac{s}{2}}m^2, \label{gm2new}
\ea and  the new cutoff function:
\be f'(k_y',k_z')=f(k_y,k_z)=f(e^{- s}k_y',e^{-\frac{3s}{2}}k_z').
\ee

We try to pick the cutoff function in a way  consistent with the spatial anisotropy of
the problem, and also for future convenience.  One reasonable choice is
\be f(k_y,k_z)= e^{-a^2 (k_z^2+ \gamma a|k_y|^3)}.
\ee The RG calculation (described in more detail in the Appendix) now proceeds as usual \cite{knopsouden} though we notice straight away that we have scaling of the temperature even before we go to
higher order terms.  Thus, even though we are looking at a 2-D problem we have
effectively higher dimensional behavior.  As is known \cite{kosterlitz77} the
Ising model is equivalent to a Coulomb gas in all dimensions and this is always
in the plasma phase for $d>2$.  Our  final results from the perturbative
renormalization group to first order in g yield the following differential renormalization equations
\ba
\frac{dg}{ds} &=&\left(\frac{5}{2}- \frac{\Gamma[\frac{2}{3}]}{8 m^2 \sqrt{\pi}
\gamma^{2/3}}\right) g,\label{dg}\\
\frac{d(m^2)}{ds}&=&\frac{1}{2} m^2\label{dm2}.
\ea The resultant flows in the $\{m^2,g\}$ plane are shown in Fig.~\ref{mgflows}.  The differential RG equations show that  $m^2$ always grows and
that the flow of the variable $g$ changes sign from positive to negative as $m^2$
becomes smaller than $ \frac{\Gamma[\frac{2}{3}]}{20 \sqrt{\pi} \gamma^{2/3}}$,
indicated by point c on the plot.  However this point is of no special importance because there is no phase transition: recalling that $m^2 \sim \frac{1}{T}$ we see
that the temperature always flows to zero.  Thus, the system remains in the same phase
with flat domain walls that it has at $T=0$,  even as the temperature becomes
finite.   This formal RG result thus confirms the earlier expectation based on
Eq.~(\ref{gshortrange}).  This result is to be contrasted with the first order flow equations for the case of short range interactions where $\frac{dm^2}{d s}=0$ and flows are vertical at this level of approximation - cf.~Eq.~(2.24) in Ref.~[\onlinecite{knopsouden}] truncated to first order in y, with y and J being the analogs of $g$ and $m^2$ respectively in our analysis.  It is interesting to note that simple tree level RG rescaling in a d-dimensional (short-range) sine-Gordon model yields $\frac{dJ}{ds}=(d-2) J$, in the analogous notation of Ref.~[\onlinecite{knopsouden}], suggesting that our model is in effectively $\frac{5}{2}$ dimensions.

In addition to this classical case, we have also considered the quantum
mechanical case, which is equivalent to a classical sine-Gordon model in one
extra dimension. Since this raises the dimensionality of the system even more,
we find that there is no quantum roughening either.


\begin{figure}[tb]
\includegraphics[width=3.375in]{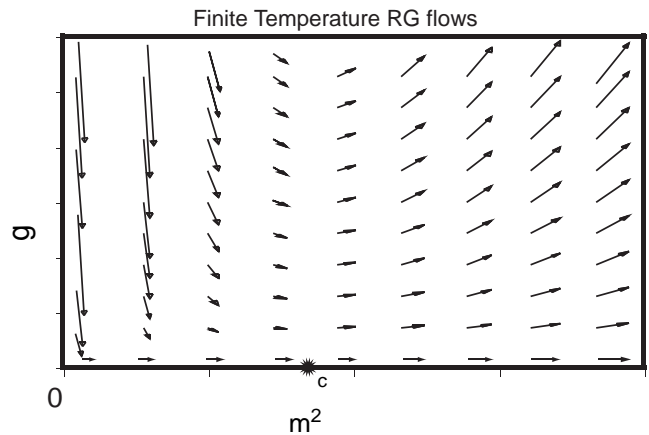}
\caption{\label{mgflows} Renormalization group flows in $\{m^2, g\}$
parameters.}
\end{figure}

\subsection{\label{steps} Further Possibilities:  Stepped Surfaces} It is also
interesting to consider the possibility of steps and investigate whether the
steps themselves are smooth or rough.  As our system we considered two
semi-infinite three-dimensional domains of antiparallel magnetization in the
z-direction and added to the domain interface a unit lattice step in the
x-direction.  The magnetization vector in this case is defined by a
one-dimensional interface height function $h(z)$,
\be
\vec{M}=\left\{ 0,0,m_0\sgn \left(x-a\Theta[y-h(z)]\right)\right\}.
\ee

For the case of short-range forces as shown in Fig.~\ref{2dphasediagram}, the
classical interface is always rough.  Here we investigate the effect of the long
range forces at finite temperatures. A similar problem has been investigated by
Thouless\cite{thouless1d} who looked at one-dimensional Ising systems that have
long range interactions $\sim r^{-2}$. He concluded that for interactions that
fall off faster than $r^{-2}$ there cannot be an ordered state, whereas for an
interactions that falls off as $r^{-2}$the order cannot go continuously to zero.
Even though our one-dimensional interface problem is a height model, not an
Ising model, the Thouless results suggest that the interface may be rough at
finite temperatures.

As before we obtain the energy for the step fluctuations in analogy with
electrostatics,
\be U_S=2 m_0^2a^2\int dz dz' \frac{h'(z)h'(z')}{\left[
\left(h(z)-h(z')\right)^2+(z-z')^2\right]^{1/2}}.
\ee

By assuming that $\theta < 1$ we get to zeroth order in
$\left(h(z)-h(z')\right)$ the step fluctuation energy in Fourier space,
\be U_{S0}=r\pi
m_0^2a^2\int\frac{dk_zdk_y}{(2\pi)^2}\frac{k_z^2}{\left(k_y^2+k_z^2%
\right)^{1/2}}|h(k_z)|^2.
\ee Our approximation scheme here introduces the necessity for an ultra-violet
cutoff, $\Lambda_y$ in the $k_y$ integral, which is set by the discreteness of
the lattice, and namely that the two height variables considered in the
correlator are not evaluated at arbitrarily close points, but are at the very
least on adjacent lattice sites.

If we now consider the mean height correlation, and once more introduce a
surface tension term, $\gamma_s k_z^2$,  we have,
\ba G_s&=&\left< \left( h(z)-h(0)\right)^2\right>\nonumber\\
&=& \int \frac{dk_z}{2\pi}\frac{1-e^{i k_z z}}{4\beta m_0^2 k_z^2\ln
\left(\frac{\Lambda_y}{|k_z|}+\gamma_s\right)}.
\ea

In the large distance limit $G_s \sim \frac{L}{\ln L}$, and the Gaussian model
is rough at all finite temperatures.  The long range interaction gives rise to a logarighmic term in the energy, raising the effective dimensionality of the system only very slightly above one.  This suggests that if we were to take into account the lattice structure by adding a periodic pinning potential to form another modified sine-Gordon model there would be no effect on the roughness of the interface since  the periodic sine term cannot pin the interface in dimensions lower than two.  This is clear if we keep in mind that the usual sine-Gordon model for short range interactions only displays a phase transition in exactly two dimensions, where the free Gaussian model has logarithmically divergent correlations. In the usual sine-Gordon model in one dimension the sinusoidal pinning potential  is unable to control the linearly diverging correlations of the free Gaussian model, and hence interfaces remain rough.   This argument suggests that for the long-range sine-Gordon model in one dimension where the effective dimensionality is close to one,  the parameters flow to the high temperature limits, and just as for the case of short range forces, the periodic potential is unable to pin the step which fluctuates freely and hence is rough
at all finite temperatures.

For the quantum case with short range forces, $G_s$ diverges as $\sim\ln L$ for
the Gaussian model and the full sine-Gordon model exhibits a roughening
transition in the KT universality class as shown in Fig.~\ref{2dphasediagram}. 
Here, for the case of long-range forces in the quantum problem, similar analysis
shows that that the divergence of $G_s\sim \sqrt{\ln L}$ is extremely weak for
the Gaussian problem, possibly indicating that the sine-Gordon model will be
smooth though we have not been able to verify this in a full RG calculation.

\section{\label{end} Discussion}
We have found that even though \Li\  interfaces initially appear to be ideal for
having a roughening transition, the same long-range interactions which account
for the system's domain structure turn out to be also responsible for the lack
of a roughening transition.  The long-range interaction term which arises from
the dipolar interactions effectively raises the dimensionality of the system and
makes it equivalent to a sine-Gordon model in dimensions greater than two. The
RG flow diagram, Fig.~\ref{mgflows}, which we obtain for $m^2\sim \frac{1}{T}$
indicates that in effect the roughening transition coincides with the bulk
order-disorder transition and the whole ferromagnetic region is smooth.  In
contrast, a step in the interface seems to have a rough profile at all
temperatures, and the long-range interactions do not raise the dimensionality of
the step enough to drive it into the smooth phase.

\begin{acknowledgments}

This work was supported by NSF DMR-0342157.  We are grateful for helpful
conversations with Subir Sachdev.
\end{acknowledgments}
\appendix*
\section{\label{details}Renormalization details} In the following we give some
more details of the smooth cutoff procedure\cite{knopsouden} leading to the differential renormalization
group equations (\ref{dg}) and (\ref{dm2}).   Beginning with our choice for a
smooth cutoff function,
\be f(k_y,k_z)= e^{-a^2 (k_z^2+ \gamma a|k_y|^3)},
\ee
and using $k_z'=e^{\frac{3}{2}s},\ k_y'=e^{s}k_y$, we obtain for small s a rescaled cutoff
\ba
 f'(k_y',k_z')&=&e^{-a^3 e^{-3 s}\left(\frac{k_z'^2}{a}+\gamma |k_y'|^3\right)}\\
&\approx&  f(k_y',k_z')+\zeta(k_y',k_z')+O(s^{2}),
\ea
with
\be
\zeta(k_y',k_z')=3 s a^3 \left(\frac{k_z'^2}{a}+\gamma |k_y'|^3\right)f(k_y',k_z').
\ee

We can now define a new field, say $\chi(\vec{r})$, having $\zeta$ as its smooth cutoff. 
We consider a sine-Gordon model for both the $\chi$ and $\phi$ fields and write
\be S'= S_0[\phi',f]+S_1[\chi,\zeta]+S_{01}[\phi',\chi],
\ee with
\ba
S_0[\phi',f]&=&  -\frac{m^2}{ \pi}  \int \frac{d^2\vec{k} e^{\frac{1}{2}s}}{(2\pi)^2 f(k_y,k_z)}
\Big{\{}\frac{k_z^2}{a |k_y|}+\nonumber\\
&\ & +\gamma (k_y^2) \Big{\}} |\phi'(\vec{k})|^2,
\ea
\ba
S_{1}[\chi,\zeta] &=& -\frac{m^2}{ \pi}  \int \frac{d^2\vec{k} e^{\frac{1}{2}s}}{(2\pi)^2 \zeta(k_y,k_z)}
\Big{\{}\frac{k_z^2}{a |k_y|}+\nonumber\\
&\ & +\gamma (k_y^2) \Big{\}} |\chi(\vec{k})|^2,
\ea
and
\be
S_{01}[\phi',\chi]=\frac{g'}{a^2} \int d^2r\cos(\phi'(\vec{r})+\chi(\vec{r})).
\ee
It is straightforward to show\cite{knopsouden} that the action $S'$ results in the same partition function corresponding to the rescaled action S in Eq.~(\ref{rescaledAction}). In Ref.~[\onlinecite{knopsouden}] this is done by mapping to the Coulomb gas, but an easier method is simply to shift the argument of the cosine term by $-\chi(\vec{r})$ and then carry out the Gaussian integral over $\chi(r)$, to recover the rescaled action S.

Now, as part of the RG calculation we integrate out the extra field $\chi$ to restore the original cutoff function.
We use a cumulant expansion, in $S_{01}$ and write
\ba Z &=&\int D\phi' e^{S_0}\left< e^{S_{01}}\right>_1\nonumber\\
&=& \int D\phi' e^{S_0}e^{\left<S_{01}\right>_1
+\frac{1}{2}\left[\left<S_{01}^{2}\right>_1
-\left<S_{01}\right>^2_1\right]+O(S_{01}^{3})},
\ea
where the averaging $<\ >_1$ indicates integrating out the extra $\chi(\vec{r})$ fields using the Gaussian action $S_1$. In addition, a multiplicative constant has been absorbed into the measure.\\

In this analysis we consider the first term in the exponential of the cumulant expansion
\ba
\left<S_{01}\right>_1 &=& \left<\int d^2 r\left(
\frac{g'}{a^2}\right)\cos(\phi'(\vec{r})+\chi(\vec{r}))\right>_1\nonumber\\
&=&\int d^2 r \left(\frac{g'}{a^2}\right)\cos(\phi'(\vec{r}))<\cos(\chi(\vec{r}))>_1-\nonumber\\
&\ &-\sin(\phi'(\vec{r}))<\sin (\chi(\vec{r}))>_1.
\ea
The average over the sine term is zero by symmetry.  For the other term we can use the result for Gaussian integrals which
states
\be
\left< e^{-i\phi(\vec{r})}\right>_{\text{Gaussian}}=
e^{-\frac{1}{2}\left<\phi(r)^2\right>_{\text{Gaussian}}},
\ee
so we obtain 
\ba
\left< S_{01}\right>_1&=&\int d^2 r \left( \frac{g'}{a^2}\right)\cos
\phi'(\vec{r})<\cos(\chi(\vec{r})>_1\nonumber\\
&=&\int d^2 r \left( \frac{g'}{a^2}\right)\cos
\phi'(\vec{r})e^{-\frac{1}{2}\left<\chi(\vec{r})^2\right>_1}.
\ea

The average of $\chi\vec({r})^2$ is given by  
\ba
\left<\chi(\vec{r})^2\right>_1&=&\int \frac{d^2 k}{(2\pi)^2} \frac{a
\pi |k_y| e^{-\frac{1}{2}s}\zeta(k_y,k_z)}{2 m^2\{  k_z^2+
\gamma a |k_y|^3\} }\nonumber\\
&=&  \frac{ s e^{- \frac{ s}{2} } \Gamma[\frac{2}{3}]}{4 m^2
\sqrt{\pi} \gamma^{2/3}}.
\ea

We gather our results to obtain
\ba
\left<S_{01}\right>_1&=&\int d^2 r \left( \frac{g'}{a^2}\right) \cos
\phi'(\vec{r})e^{-\frac{ s e^{- \frac{ s}{2} }
\Gamma[\frac{2}{3}]}{8 m^2 \sqrt{\pi} \gamma^{2/3}} }\nonumber\\
&\approx& \left( \frac{g_s }{a^2}\right)\int d^2 r \cos \phi'(\vec{r})+O(s^{2}),
\ea
where we have introduced the renormalized coupling $g_s$, which is related to
the original $g$ coupling:
\ba g_s&=&(1- \frac{ s \Gamma[\frac{2}{3}]}{8 m^2 \sqrt{\pi}
\gamma^{2/3}})g'\nonumber\\
&\approx&\left( 1+\left(\frac{5}{2}- \frac{\Gamma[\frac{2}{3}]}{8 m^2 \sqrt{\pi}
\gamma^{2/3}}\right) s \right)g.
\ea

From the above relation we obtain the differential renormalization equation (\ref{dg}).  Note that $m^{2}$ receives no
further corrections to this order in the cumulant expansion and hence Eq.~(\ref{dm2}) is obtained directly from Eq.~(\ref{gm2new}).


\end{document}